\documentclass[a4paper,11pt]{article}
\usepackage{cite}

\usepackage[margin=1.2in]{geometry}

\usepackage{times}
\usepackage{graphicx}
\usepackage{bm}
\usepackage{multirow}
\usepackage{braket}

\usepackage{amssymb}
\usepackage{amsfonts}
\usepackage{amsmath}
\usepackage{graphicx}
\usepackage{bm}
\usepackage{multirow}
\usepackage{cite}

\usepackage{etoolbox}
\apptocmd{\thebibliography}{\setlength{\itemsep}{0pt}}{}{}

\newcommand{\op}{\hat}
\newcommand{\hbn}{hBN}

\usepackage{fancyhdr}

\newcommand{\vect}{\boldsymbol}
\newcommand{\colvect}[2] {\begin{pmatrix} #1 \\ #2\end{pmatrix}}

\begin{document}

  \title{Moir\'e superlattice effects in graphene/boron-nitride van der Waals heterostructures}
\author{J.~R.~Wallbank$^1${\footnote{e-mail: j.wallbank@lancaster.ac.uk}}, M.~Mucha-Kruczy\'{n}ski$^2$, Xi Chen$^1$,  V.~ I.~Fal'ko$^1$} 
\date{}

 \maketitle
 
 \thispagestyle{fancy}
\fancyhead{ \Large{\bf{Review Article} } }
\rhead{}
\chead{}

\noindent{\small \textit{ 
$^1$Physics Department, Lancaster University, LA1 4YB, UK\\
$^2$Department of Physics, University of Bath, Claverton Down, Bath, BA2 7AY, UK
} }

   \begin{abstract}
   Van der Waals heterostructures of graphene and hexagonal boron nitride feature a moir\'e superlattice for graphene's Dirac electrons.
Here, we review the effects generated by this superlattice, including a specific miniband structure featuring gaps 
and secondary Dirac points, and a fractal spectrum of magnetic minibands known as Hofstadter's butterfly.
\end{abstract}
% \tableofcontents
 
\section{Introduction}

Van der Waals heterostructures formed from stacks of two-dimensional crystals have recently attracted a great deal of interest, as the available materials cover a broad spectrum of physical properties \cite{novoselov_phys_scr_2012,bonaccorso_mat_2012,geim_nat_perspective_2013,yankowitz_jpcm_2014}. The first and, by now, the most developed of these heterostructures is that formed by graphene placed on hexagonal boron nitride ({\hbn}). Both materials are made from a single layer of atoms arranged in a honeycomb structure. However, whereas graphene \cite{Castro_Neto_RevModPhys_2009,novoselov_nature_2012} is a gapless semiconductor, {\hbn} is an insulator with a band gap of approximately $6\,$eV \cite{watanabe_naturemat_2004}. For this reason {\hbn} was first used as a substrate to preserve graphene's electronic properties \cite{dean_naturenano_2010}. Nevertheless, the small difference between the lattice constants of the two crystals, 
and any misalignment between their respective crystallographic axes, generates a large-scale quasi-periodic hexagonal pattern, known as the moir\'e superlattice (mSL) \cite{xue_natmat_2011,decker_nanolett_2011,yankowitz_natphys_2012}. In this review, we focus on the qualitatively new phenomena generated by the presence of the mSL, such as the formation of electronic and magnetic miniband structure. Since the perturbing effect of the mSL on graphene's Dirac electrons decreases with increasing misalignment between the {\hbn} and graphene lattices, we concentrate on the well-aligned heterostructures (misalignment angle $\theta \lesssim 2^\circ$).

The mSL provides a long-wavelength periodic potential, which leads to the Bragg scattering of electrons in graphene. This generates a miniband structure with the most pronounced effects found at the edges of the first and second miniband, where the mSL perturbation leads to the generation of secondary Dirac Points (sDPs). The signature of these sDPs has been observed, both in the density of states \cite{yankowitz_natphys_2012,yu_natphys_2014} and in transport measurements \cite{ponomarenko_nature_2013,  hunt_science_2013, yang_naturemat_2013}, as a repetition of the behaviour of the primary Dirac point when the chemical potential is tuned to the energy of the sDP. Notably, the signatures of the sDPs are much more pronounced in the valence band than the conduction band, which is a manifestation of the electron-hole symmetry breaking produced by the mSL perturbation.

Further spectral reconstruction occurs at the neutrality point of graphene. While the first transport measurement \cite{ponomarenko_nature_2013} reported a gapless spectrum, another experiment \cite{hunt_science_2013} reported a band gap of approximately $30\,$meV. The exact details of the spectrum at the neutrality point appear to depend on the architecture of the device, namely whether the graphene is encapsulated from the top with an additional, intentionally misaligned, {\hbn} layer, as was the case in the former experiment, or left unencapsulated, as it was in the latter.
A further study \cite{woods_natphys_2014} showed that the graphene and hBN layers in almost perfectly aligned, yet unencapsulated devices, undergo local displacements subject to their mutual adhesion, so that they assume their preferred stacking order for a large portion of the moir\'e unit cell. In these devices transport measurements revealed significant band gaps, consistent with the expectation \cite{san-jose_prb_2014,san-jose_prb_2014_2,bokdam_prb_2014,jung_arxiv_2014} that graphene's sublattice symmetry is strongly broken in this phase. Conversely, the lack of observed band gap, and absence of strain signatures in the Raman spectroscopy data, was interpreted as evidence that the additional randomised adhesion forces due to encapsulation with a misaligned {\hbn} layer suppress the local displacements of the crystals, and therefore (on average) restore the sublattice symmetry. 
Also, recent magneto-optical spectroscopy data \cite{chenZG_natcom_2014} for unencapsulated epitaxially grown heterostructures has been fitted to a gapped Dirac spectrum with a gap of $38\,$meV and a transition dependent Fermi velocity, the latter interpreted as a signature of the electron-electron interaction in the graphene/hBN heterostructure \cite{song_prl_2013,bokdam_prb_2014,jung_arxiv_2014}.

When the graphene/hBN heterostructure is placed in a magnetic field that is strong enough that the magnetic length is comparable to the periodicity of the mSL, an intricate fractal spectrum of magnetic bands \cite{ponomarenko_nature_2013, dean_nature_2013, hunt_science_2013, yu_natphys_2014} (sometimes referred to as ``Hofstadter's butterfly'' \cite{hofstadter_prb_1976}) is formed. The physics of this phenomena was explored theoretically many years ago \cite{brown_physrev_1964, zak_physrev_1964}. However, accessing the fractal electronic spectrum experimentally is more challenging because, for a typical crystalline periodicity, the magnetic fields required to access this regime are unobtainable. Low-field oscillations  and internal structure within Landau 
levels  \cite{weiss_europhys_1989,weiss_prl_1991,pfannkuche_prb_1992,ferry_pqe_1992,albrecht_prl_1999,schlosser_prl_2001,albrecht_prl_2001,geisler_prl_2004} have been observed using semiconductor heterostructures, but the technological challenges involved in producing the structures and accessing the two-dimensional electron gas limited further studies. 
In contrast, graphene/{\hbn} heterostructures are uniquely suited to investigate the ``Hofstadter butterfly'' spectrum because of their high electronic quality, ease of doping, and convenient range of the moir\'e wavelength ($A\lesssim14$nm). One striking feature in these heterostructures is the systematic reappearance of tertiary Dirac points at the edges of the magnetic minibands formed at rational values of the magnetic flux \cite{ponomarenko_nature_2013,  chen_prb_2014,diez_prb_2014}.

Graphene/{\hbn} heterostructures have also been studied using other techniques, such as optical absorption \cite{shi_natphys_2014} and Raman \cite{eckmann_nanolett_2013} spectroscopy. For the former, the reconstruction of the electronic spectrum at the edge of the first miniband and the modification of optical selection rules, result in a modulation \cite{abergel_njp_2013, shi_natphys_2014} of the otherwise flat absorption spectrum of graphene \cite{nair_science_2008}. In the case of the latter, the width of the Raman 2D peak (for unencapsulated samples) was found to increase linearly with moir\'e wavelength, which was interpreted in terms of increasing strain in the graphene layer \cite{eckmann_nanolett_2013}.
It has also been calculated that the formation of minibands in the graphene/{\hbn} heterostructures should impact graphene's plasmon dispersion \cite{tomadin_prb_2014}.
Finally, the moir\'e minibands in graphene/{\hbn} heterostructures may have some interesting topological properties including non-zero Berry curvature \cite{song_arxiv_2014,san-jose_prb_2014_2}. In an 
applied electric field the Berry curvature generates an 'anomalous' contribution to the electron's velocity \cite{xiao_revmodphys_2010}, which has been observed in recent transport measurements \cite{gorbachev_science_2014}.
 
Although not discussed in detail here, it is important to note that the electronic quality of graphene-based devices is dramatically improved by using {\hbn} as a substrate \cite{dean_naturenano_2010, xue_natmat_2011,decker_nanolett_2011,mayorov_nanolett_2011,gannett_apl_2011,kim_apl_2011,wang_ieee_2011,bresnehan_acsnano_2012,dean_ssc_2012}, regardless of the misalignment between the two crystals. As compared to the traditional SiO$_2$ substrate \cite{martin_natphys_2008, cho_prb_2008, zhang_nat_phys_2009}, or other tested substrates  \cite{tan_apl_2014, kretinin_nanolett_2014}, {\hbn} is flatter, contains less charged impurities, and forms a ``self-cleaning'' interface with graphene \cite{haigh_nat_mat_2012,kretinin_nanolett_2014}. All of these factors contribute to an increased mobility of electrons in graphene and decreased charge inhomogeneities due to electron-hole puddles.   With improvements in the manufacturing 
techniques used to create 
these heterostructures \cite{reina_jpcc_2008,zomer_apl_2011,wang_science_2013,zomer_arxiv_2014}, as well as developments in the growth of graphene directly on top of the {\hbn} layer \cite{bjelkevig_jpcm_2010,usachov_prb_2010,son_nanoscate_2011,ding_carbon_2011,tang_carbon_2012,garcia_ssc_2012,kim_nano_lett_2013,lui_nanolett_2013,roth_nano_lett_2013,yang_naturemat_2013,tang_sci_rep_2013}, {\hbn} looks set to be the substrate of choice for graphene devices for many years to come.\newline

\begin{figure}[tbhp]
\centering
\includegraphics[width=.5 \textwidth]{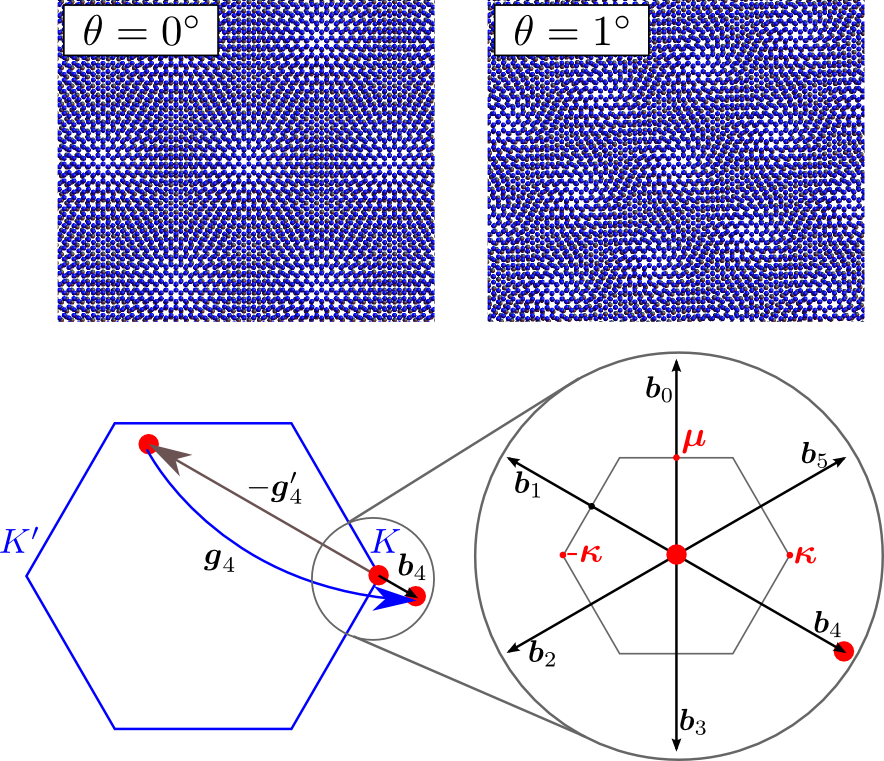}
\caption{Top: examples of the moir\'e pattern arising from graphene (blue) placed on {\hbn} (grey), for two different misalignment angles, $\theta=0$ and $\theta=1^{\circ}$. Bottom: the Brillouin zones of graphene (left) and the mSL (right), with various high symmetry points and reciprocal lattice vectors labelled.
}
\label{fig:moire_geom}
\end {figure}

\emph{Geometry of the moir\'e superlattice.}
When two isostructural crystals with a small lattice mismatch, $\delta$, are placed on top of each other with a small misalignment angle, $\theta$, between their respective crystal directions, the beating of the two crystalline periodicities produces the large-scale modulation of their local stacking known as the mSL (see Fig.~\ref{fig:moire_geom} upper panels). For graphene, this geometry-based effect occurs on a variety of different hexagonal substrates/surfaces \cite{rong_prb_1993, berger_jpcb_2004, diaye_prl_2006, marchini_prb_2007, sutter_prb_2009, gao_nanolett_2010, sicot_ascnano_2012, wintterlin_surfsci_2009, batzill_ssr_2012}, and can be observed, for example, using scanning tunnelling microscopy or conducive atomic force microscopy 
\cite{hunt_science_2013,ponomarenko_nature_2013,yankowitz_natphys_2012, dean_nature_2013, xue_natmat_2011, decker_nanolett_2011, roth_nano_lett_2013, yang_naturemat_2013, woods_natphys_2014, tang_sci_rep_2013, yankowitz_jpcm_2014}. In Sec.~\ref{sec:moire_magnification}, we discus the magnifying effect in which defects in either of the two crystal 
layers (such as dislocations or wrinkles) generate a large partner defect in the mSL \cite{coraux_nano_lett_2008,hermann_jpcm_2012,cosma_faraday_2014}. Here, we assume that graphene and its substrate are perfect crystals that are placed rigidly on top of each other. Then the six ($m=0,\cdots5$) shortest reciprocal vectors of the mSL,
\begin{align*}%\label{eq1:moire_harmonic}
 \vect b_m=\vect g_m-\vect g'_m\approx \delta \vect g_m -\theta \vect l_z\times \vect g_m,
\end{align*}
are obtained from the graphene reciprocal lattice vectors $\vect g_m=\op{R}_{2\pi m/6}(0,\frac{4\pi}{\sqrt{3}  a})$ by the subtraction of the substrate reciprocal lattice vector $\vect g_m'=(1+\delta)^{-1}\op{R}_\theta \vect g_m$ (where $\op{R}_\theta$ is the matrix for anticlockwise rotation, $a$ is the graphene lattice constant, and $\theta,\delta\ll1$). The corresponding real space lattice vectors of the mSL are given by,
\begin{align}\label{eq:hatM}
\vect A_m=\hat M\vect a_m,\quad \hat M\equiv\frac{1}{\delta^2+\theta^2}\begin{pmatrix} \delta&\theta\\-\theta&\delta \end{pmatrix},
\end{align}  
where $\vect a_m=\hat R_{2\pi m/6}(a,0)$ are the graphene lattice vectors. It should be noticed that the mSL described by vectors $\vect A_m$ is distinct from the crystallographic lattice of the two layer-system, the latter requiring a perfect translational symmetry that is only present when the two layers are commensurate.  

As shown in the top row of Fig.~\ref{fig:moire_geom}, as $\theta$ increases, the primitive lattice vectors of the mSL are rotated by $\Theta\approx-\tan^{-1}(\theta/\delta)$ and their length decrease as $A\equiv |\vect A_m|\approx(\delta^2+\theta^2)^{-1/2}a$. For the graphene/{\hbn} heterostructure, $\delta\approx 1.8\%$ \cite{liu_prb_2003} so that mSLs with lattice constants as large as $A\approx14\,$nm are possible for $\theta=0$. The energy of the first miniband edge is approximately $v b/2\approx v(\delta^2+\theta^2)^{1/2}|\vect g_m|/2$, which is as low as $170\,$meV for $\theta=0$ (where $b\equiv|\vect b_m|$, $\hbar=1$, and $v$ is the Dirac velocity). Importantly, this energy corresponds to a carrier density of $n\sim 2\times10^{12}\,\text{cm}^{-2}$, so that electrostatic doping can be used to shift the Fermi level to the miniband edge, and the strong spectral reconstructions found at the edge of the superlattice Brillouin zone (sBZ) can be probed in transport measurements. To compare, the 
characteristic 
energy $vb/2$ increases rapidly with the misalignment angle, so that for $\theta=2^\circ$ almost five times greater doping is required to fill the first miniband.\newline

\emph{Superlattice perturbation for Dirac electrons.} 
To model the low-energy band structure ($|\epsilon|\lesssim v b$) of the graphene/{\hbn} heterostructure, we use the fact that {\hbn} has a large band gap ($\Delta_{  \text{\hbn} }\sim6\,$eV \cite{watanabe_naturemat_2004}) , to project the full two-layer Hamiltonian onto the Hilbert space of the graphene $\pi$ orbitals only.
Then, the dominant effect of the substrate on graphene Dirac electrons can be modelled by scattering processes using the six shortest non-zero reciprocal lattice vectors of the {\hbn} substrate (sometimes referred to as the ``first star'' of reciprocal lattice vectors).  
One such process, in which an electron (red point) at graphene's $K$ Brillouin zone corner is scattered by $-\vect g'_4$, is displayed in the left lower panel of Fig.~\ref{fig:moire_geom}. Hence, upon addition of a graphene reciprocal vector, the mSL perturbation provides intravalley scattering by the simplest harmonics of the mSL ($\vect b_4=\vect g_4-\vect g'_4$ in this case). 
Scattering processes involving higher reciprocal lattice vectors of the substrate (or, equivalently, of the mSL) are suppressed, as they necessarily depend on the overlap between higher Fourier components of the graphene $2p_z$ orbital (see section \ref{sec:microscopic_models}). However, at interlayer distances, $z$, comparable to the graphene-{\hbn} interlayer spacing, $d$, these Fourier components decay rapidly as a function of the in-plane momentum $\vect{q}$,
\begin{align}\label{eq:pz_FT}
\hat\psi(\vect q,z)=\frac{  z\,(2a_0+\sqrt{X}|z|)  }{2\sqrt{2\pi} (a_0 X)^{3/2}}e^{-\sqrt{X}|z|/(2a_0)},
 \qquad X\equiv1+4a_0^2|\vect q|^2,
\end{align}
where $\hat\psi(\vect q,z)$ is the two-dimensional in-plane Fourier transform of the graphene $2p_z$ orbital, and $a_0$ is the effective carbon Bohr radius.

\begin{figure*}[tbhp]
\centering
\includegraphics[width=.9\textwidth]{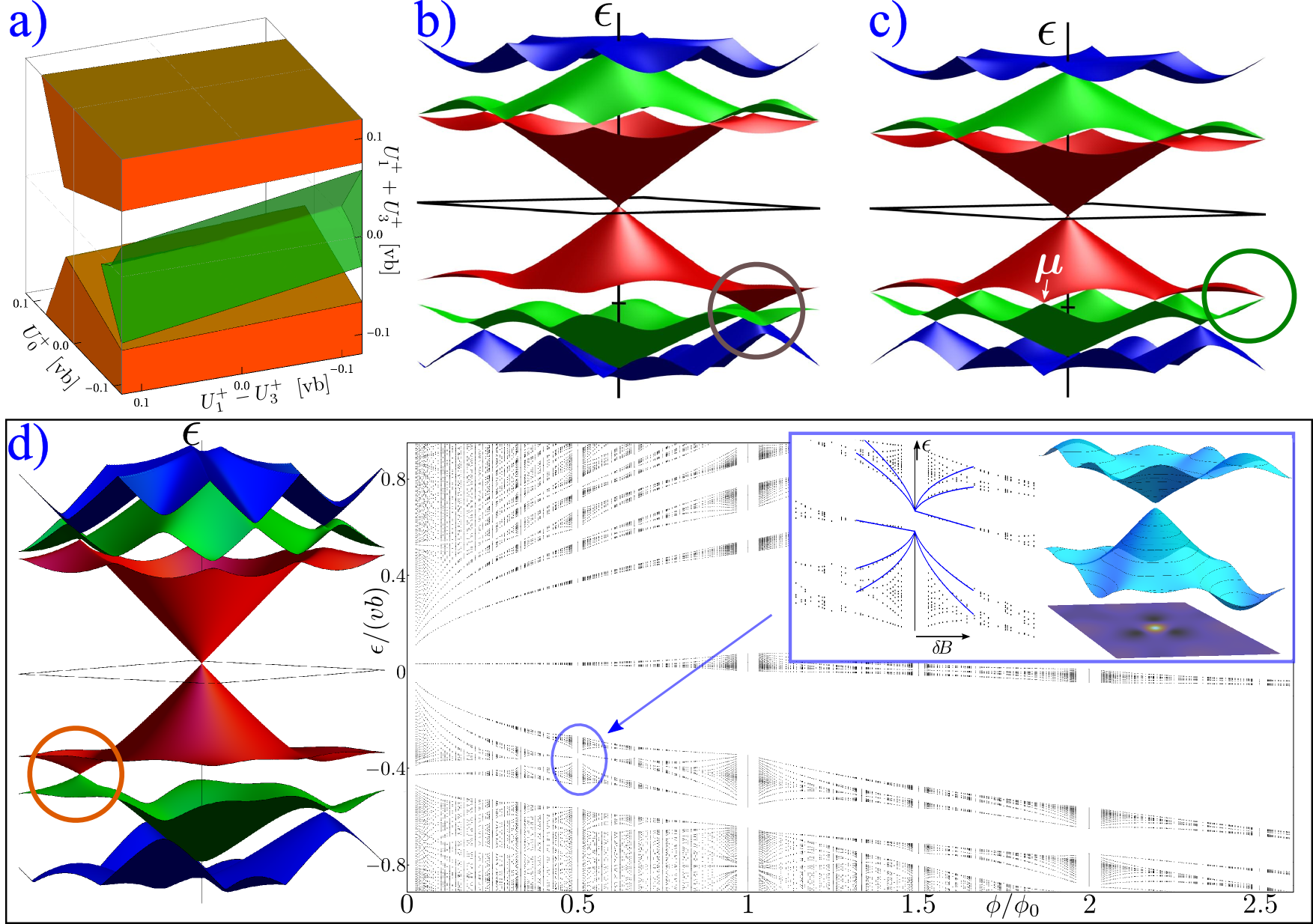}
\caption{
(a) Three volumes in the space of the inversion-symmetric mSL parameters where the edge of the first miniband, in graphene's valence band, contains a single isolated sDP (orange), three isolated sDPs (green), or overlapping bands (white).  
Representative band structures are shown for each regime, calculated for (d) $U^+_0\!=\!0.032vb$, $U^+_1\!=\!-0.064vb$, $U^+_3\!=\!-0.055vb$, (c) $U_0^+\!=\!-0.15vb$,  or (b) $U_1^+\!=\!U_3^+\!=\!0.075vb$. 
The fractal spectrum of magnetic minibands
and an example showing the full dispersion of a magnetic miniband are also displayed for the realisation of mSL potential in (d).
}
\label{fig:parameter_space}
\end{figure*}

In addition to the dominance of the simplest mSL harmonics in the perturbation, the {\hbn} substrate can only affect the graphene electrons via three distinct mechanisms \cite{wallbank_prb_2013}; (i) an electrostatic potential, which does not distinguish between the two carbon sublattices, (ii) a sublattice-asymmetric part of the potential, and (iii) spatial modulation of the nearest neighbour carbon-carbon hopping amplitude. Each of these can be thought of as made of two contributions, either symmetric or antisymmetric under the in-plane spatial inversion. It can be argued \cite{wallbank_prb_2013} that in the two limits where either, both boron and nitrogen sublattices perturb the Dirac electrons with almost the same strength, or, the dominant perturbation arises from one sublattice only, the inversion symmetry of the system would only be weakly broken. Then, the mSL potential can be modelled as a combination of a dominant inversion-symmetric part with the addition of a small inversion-asymmetric 
perturbation. Accordingly, the mSL 
perturbation can be parametrised by 
three phenomenological parameters, $U^+_0$, $U^+_3$ and $U^+_1$, each controlling the strength of the inversion symmetric component of modulation mechanisms (i), (ii) and (iii) described above. For such perturbations, the electronic miniband spectra in either graphene's valence or conduction bands can be classified into three groups depending on the mutual arrangement of the first and second miniband: either (a) they do not overlap and are connected by a single isotropic sDP, (b) they do not overlap and are connected by a triplet of anisotropic sDPs, or (c) the minibands overlap. The regions in the parameter space of the three inversion-symmetric parameters, corresponding to the valence band spectrum of a particular type, are shown in different colours (orange, green and white, respectively) in Fig. \ref{fig:parameter_space}(a). Representative examples of spectra from each region are shown in Fig. \ref{fig:parameter_space} (d,c,b). Note, that while a purely inversion-symmetric mSL perturbation generates 
both gapless primary 
DP and sDPs, the addition of a finite inversion-asymmetric component will open gaps at both. 
The experimental evidence \cite{yankowitz_natphys_2012,ponomarenko_nature_2013,  hunt_science_2013, yang_naturemat_2013, yu_natphys_2014} clearly points to the existence of sDPs in the valence band, with recent temperature-dependent measurements attributed to the spectrum featuring only a single isotropic sDP \cite{yu_natphys_2014}.

In a perpendicular magnetic field, a hierarchy of magnetic minibands and gaps is generated for each magnetic field providing a rational fraction of the flux quantum $\phi\!=\!\phi_0p/q$ (here $\phi_0=h/e$ is the flux quantum) per moir\'e unit cell of the graphene/{\hbn} heterostructure. 
For a weak magnetic field ($\phi \lesssim 0.2 \phi_0$) these can be traced to the weakly broadened Landau levels of the zero-magnetic field bandstructure (see Fig.~\ref{fig:parameter_space} d). 
In contrast, in stronger magnetic fields ($\phi\gtrsim0.2\phi_0$), the edges of many of the consecutive magnetic bands can be described in terms of a weakly gapped Dirac spectrum \cite{ponomarenko_nature_2013,chen_prb_2014}. One example is shown as an inset in Fig.~\ref{fig:parameter_space}(d) with the Berry curvature of the lower magnetic miniband shown as a colour map. The fractal spectrum surrounding each magnetic miniband can be interpreted \cite{chang_prl_1995} in terms of the weakly broadened Landau levels of these next generation Dirac electrons (see right panel of inset). This can be seen best in the cases when the gap between consecutive minibands is small, and the largest gaps are around the broadened ``$n\!=\!0$'' Landau level of the corresponding effective Dirac model. 
 
Finally, we note that superlattice perturbations for Dirac electrons have been discussed in a variety of different contexts, some of which pre-date the realisation of the graphene/{\hbn} heterostructure. 
Many theoretical works have investigated the influence of one or two-dimensional electrostatic potentials on graphene electrons  \cite{park_natphys_2008, barbier_prb_2008, pedersen_prl_2008, park_nanolett_2008, park_prl_2008, abedpour_prb_2009, tiwari_prb_2009, park_prl_2009, brey_prl_2009, snyman_prb_2009, barbier_prb_2009, barbier_prb_2010, wang_prb_2010, sun_prl_2010, gattenlohner_prb_2010, guinea_ptrsa_2010, arovas_njp_2010, burset_prb_2011, kiselev_prb_2011, wu_prb_2012, pellegrino_prb_2012, esmailpour_physicae_2013, wang_pra_2014}, with the former situation realisable using patterned gates \cite{dubey_nano_lett_2013,drienovsky_prb_2014}.
Magnetic and pseudo-magnetic field superlattices (the latter arising from periodically strained graphene) have also been extensively studied \cite{snyman_prb_2009, dellanna_prb_2009, guinea_nat_phys_2010, taillefumier_prb_2011, neek-amal_apl_2014_2}, with steps towards experimental realisation \cite{tomori_ape_2011,reserbat-plantey_nanolett_2014}.
There has also been significant work on the aligned heterostructures of bilayer graphene with hBN, including the observation of Hofstadter's butterfly in transport measurements \cite{dean_nature_2013}. Theoretically, the mSL perturbation of this heterostructure can be modelled  in a similar manner to the monolayer-graphene/hBN heterostructure, except that the perturbation is felt much more strongly by the graphene layer which is closest to the hBN \cite{mucha-kruczynski_prb_2013, moon_prb_2014}. Because of this, the inversion symmetry of this heterostructure is broken, typically leading to gaps both at zero energy and at the edge of the first miniband \cite{mucha-kruczynski_prb_2013}. In a magnetic field, the broken inversion symmetry manifests itself as a strongly broken valley symmetry in the magnetic miniband structure \cite{moon_prb_2014}. 
Other systems in which miniband structure is generated by a mSL include twisted bilayer graphene \cite{santos_prl_2007, bistritzer_prb_2010, bistritzer_pnas_2011, santos_prb_2012,moon_prb_2013, bistritzer_prb_2011, moon_prb_2012, wang_nano_lett_2012, hasegawa_prb_2013, moon_prbR_2013}, graphene with almost commensurate $\sqrt{3}\times\sqrt{3}$ hexagonal crystals \cite{wallbank_prb_2013b}, and graphene on metal catalysts such as Ir(111) \cite{pletikosic_prl_2009}.

\section{Graphene on aligned {\hbn}}
\subsection{Phenomenological Hamiltonian}

To describe the effect of the moir\'e perturbation on the graphene electrons, we assume that the local form of the Hamiltonian is determined by the local relative displacement,
\begin{equation}
\vect{u}(\vect{r})=\delta\vect{r}+\theta\vect l_z\!\times\!\vect{r}+\vect{u}^{\text{\hbn}}(\vect{r})-\vect{u}^{\text{Gr}}(\vect{r}),
\end{equation}
of atoms in the underlay with respect to the graphene layer. The first two terms in $\vect{u}(\vect{r})$ describe the effect of a finite lattice mismatch, $\delta$, and misalignment angle, $\theta$, between the two lattices. The final two terms account for additional displacements in either the {\hbn} layer ($\vect{u}^{\text{\hbn}} $) or the graphene layer ($\vect{u}^{\text{Gr}}$) which may be caused, e.g., by a defect in one of the crystal layers, or spontaneous lattice distortion produced by their mutual adhesion. We also recognize that the interlayer separation between graphene and {\hbn} may vary locally, by a small amount $\Delta\!z$, for regions of the mSL with different local values of $\vect u(\vect r)$. Below, this is included as an overall prefactor to the mSL potential $h(\vect u(\vect r))\sim 1 +O(\Delta\!z)$. Then, consistent with the dominance of the first star of moir\'e harmonics in the mSL potential, a general \cite{cosma_faraday_2014} 
phenomenological Hamiltonian of the graphene/{\hbn} heterostructure reads, 

\begin{align}\label{eq:H}
&\op{H}(\vect r)=v\left(\vect{p} +e\zeta \vect{\mathcal{A}} \right)\!\cdot \!\vect{\sigma}+  \frac{a}{2}\frac{\partial\epsilon_{2p}}{\partial a} \nabla\! \cdot \!\vect{u}^{\text{Gr}}(\vect{r})+
h(\vect u(\vect r) ) \left[\delta\!\op{H}_+(\vect u(\vect r) )+\delta\!\op{H}_-(\vect u(\vect r) )\right],\\
&\delta\!\op{H}_{\pm}=\sum_{m} \left[(\pm1)^{(m+\frac{1}{2})} U^\pm_0 + \zeta (\mp1)^{(m+\frac{1}{2})}U_3^\pm \sigma_3 -\zeta i  (\mp1)^{m+\frac{1}{2}}U_1^\pm  \frac{\vect a_m}{a}\!\cdot\!\vect{\sigma}  \right]  e^{i \vect{g}_m\cdot\vect u( \vect{r})}.\nonumber
\end{align}
 Above, the Pauli matrices $\sigma_{i=1,2,3}$ and $\vect{\sigma}=(\sigma_1,\sigma_2)$ act on the electron amplitudes on the graphene $A$ and $B$ sublattices and $\zeta=+1$ ($\zeta=-1$) is used for graphene's $K$ ($K'$) valley. The first term in $\op{H}$ is the Dirac Hamiltonian, with momentum $\vect{p}=-i\nabla+e\vect{A}$ taking into account the magnetic vector potential $\vect{A}$ related to the external magnetic field using $B=\nabla\times\vect A$. 
In this term we also include one of the two dominant strain-induced effects on graphene electrons \cite{suzuura_prb_2002, manes_prb_2007, Castro_Neto_RevModPhys_2009, mucha-kruczynski_ssc_2012, manes_prb_2013} using the pseudo-vector potential $\vect{\mathcal{A}}=\frac{-\sqrt{3} \eta_0}{2 e a}(\partial_yu^\text{Gr}_y-\partial_xu^\text{Gr}_x ,\partial_xu^\text{Gr}_y + \partial_yu^\text{Gr}_x )^T$ (with $\eta_0=\partial\text{ln}\gamma_0/\partial\text{ln} a $ used to express the change in the nearest neighbour hoping $\gamma_0$ with the change of the lattice constant, and $\vect{u}^{\text{Gr}}=(u^\text{Gr}_{x},u^\text{Gr}_{y})$). 
The second term describes the other strain-induced effect: a scalar potential accounting for modification of the on-site energy $\epsilon_{2p}$ of the graphene $2p_z$ orbitals due to the local changes of bond lengths.
The remaining terms in $\hat H$ describe the mSL perturbation, where $\delta\!\op{H}_{+}$ ($\delta\!\op{H}_{-}$) contains terms symmetric (anti-symmetric) under the in-plane spatial inversion. These two contributions are written using two sets of phenomenological parameters, $U_i^{+}$ and $U_i^{-}$, which characterise the various perturbation mechanisms described in the introduction.

When the atomic rearrangements are suppressed, $\vect u^{\text{Gr}}=\vect u^{\text{hBN}}= { \Delta\!{z}}=0$, e.g.~by encapsulation of the graphene layer, the Hamiltonian is simplified to \cite{wallbank_prb_2013},
\begin{equation}\label{eq:H_simple}
\op{H}(\vect r)=v\vect{p}\cdot \vect{\sigma}+
\sum_{m} \left[(\pm1)^{(m+\frac{1}{2})} U^\pm_0 + \zeta (\mp1)^{(m+\frac{1}{2})}U_3^\pm \sigma_3 -\zeta i  (\mp1)^{m+\frac{1}{2}}U_1^\pm  \frac{\vect a_m}{a}\!\cdot\!\vect{\sigma}  \right]  e^{i \vect{b}_m\cdot\vect{r}}.
\end{equation}
Indeed, since the higher harmonics of the mSL, induced by either $\vect u^{\text{Gr}}$, $\vect u^{\text{hBN}}$, or ${ \Delta\!{z}}\neq0$ in Eq~\eqref{eq:H}, only lead to second order corrections to the energy of the lowest energy minibands, they can often be neglected.
Then Hamiltonian \eqref{eq:H} can be re-parametrised as Hamiltonian \eqref{eq:H_simple}, albeit with two additional non-spatially dependent terms \cite{san-jose_prb_2014_2}: (i) a mass term $\zeta m\sigma_3$ which opens a gap in the primary Dirac point and (ii) a trivial term which causes a shift of the energy scale. For this reason, unless stated otherwise, we limit ourselves to the discussion of Eq.~\eqref{eq:H_simple}. Also, for convenience \cite{wallbank_prb_2013} we usually choose $\theta=0$.
 
The bandstructure of Hamiltonian \eqref{eq:H_simple} calculated in either valley obeys the $c_{3v}$ symmetry. Moreover, if either the time-reversal (i.e. $B=0$) or spatial inversion symmetry (i.e.  $U^-_i=0$) is satisfied, the spectrum of Hamiltonian \eqref{eq:H_simple} obeys the symmetry $\epsilon_{\vect K+\vect p}=\epsilon_{\vect K'-\vect p}$ so that we can limit the discussion of minibands to the $K$ valley. Moreover, using the commutation properties of $\sigma_i$, the following symmetry relations for the spectrum can be obtained, 
\begin{align*}%\label{eq:sym}
\!\epsilon^{U^\pm_0,U^\pm_1,U^\pm_3}_{\vect K+\vect p}
\!=\!-\epsilon^{-U^\pm_0,U^\pm_1,-\!U^\pm_3}_{\vect K+\vect p}
\!=\!-\epsilon^{\mp U^\pm_0,\mp\!U^\pm_1,\pm U^\pm_3}_{\vect K-\vect p}.
\end{align*}
The first equality above allows one to obtain the phase diagram for the behaviour of the first miniband edge in the conduction band from that of the valence band (Fig.~\ref{fig:parameter_space}(a)). Finally, a simultaneous transformation of all the perturbation parameters \cite{wallbank_thesis_2014},
\begin{align}
\begin{pmatrix} U_0^+ \\  U_0^-\end{pmatrix}\rightarrow \op{R}_{- \frac{2\pi}{3}}\!\begin{pmatrix}U_0^+ \\  U_0^-\end{pmatrix},\quad
\begin{pmatrix} U_1^+ \\  U_1^-\end{pmatrix}\rightarrow \op{R}_{ \frac{2\pi}{3}}\!\begin{pmatrix}U_1^+ \\  U_1^-\end{pmatrix},\quad
\begin{pmatrix} U_3^+ \\  U_3^-\end{pmatrix}\rightarrow \op{R}_{ \frac{2\pi}{3}}\!\begin{pmatrix}U_3^+ \\  U_3^-\end{pmatrix},\label{eq:rotate}
\end{align}
is equivalent to a coordinate shift $H(\vect r )\rightarrow H(\vect r- \frac{4\pi}{3b^2}\vect b_0)$ and therefore leaves the band structure unchanged. 

\subsection{Moir\'e minibands for undeformed heterostructures}

Figure \ref{fig:parameter_space} shows representative spectra calculated perturbatively \cite{wallbank_prb_2013} for inversion-symmetric mSL perturbations in a basis of plane wave states from graphene's $K$ valley. All of these spectra show a gapless Dirac spectrum persisting near the conduction-valence band edge with an almost unchanged Dirac velocity. In contrast, the inversion-asymmetric terms $U^-_{i}$ are able\cite{kindermann_prb_2012} to open a minigap, $\Delta_0$,  at the primary Dirac point  \cite{mucha-kruczynski_prb_2013, abergel_njp_2013},  
\begin{align*}
\Delta_0 =\frac{24}{vb}|U^+_{1}U^-_{0}+U^+_{0}U^-_{1}|.
\end{align*}
More sophisticated calculation shows that this gap may be enhanced by either $\vect u_d\neq0$, $\Delta\!z\neq 0$, or the electron-electron interaction \cite{song_prl_2013, bokdam_prb_2014, jung_arxiv_2014,san-jose_prb_2014, san-jose_prb_2014_2}.

For the point $\vect \mu=\vect b_0/2$ on the edge of the first sBZ, zone folding brings together two degenerate plane wave states, $|\vect \mu+\vect q\rangle$ and $|\vect \mu+\vect b_3+\vect q\rangle$. Splitting of these degenerate states by the moir\'e potential in Eq.~\eqref{eq:H_simple} can be studied using degenerate perturbation theory. The corresponding $2\times 2$ matrix, expanded in small deviation $\vect q$ of the electron momentum from the $\vect \mu$-points reads,
\begin{align}
&\op{H}_{\vect\mu+\vect q}= \begin{pmatrix}
E_{\vect \mu} +s v q_y & H_{12}\\
H_{12}^* & E_{\vect \mu}-sv  q_y \\           
\end{pmatrix}, \label{eq:mu}\\
 &E_{\vect \mu}\approx svb \left(\frac{1}{2}+\frac{ q_x^2}{b^2}  \right)  ,\nonumber \\
&H_{12}\approx   (sU^+_1-U^+_3)-i(sU^-_1- U^-_3)+2 \frac{q_x}{b} (U^+_0+i U^-_0). \nonumber  
\end{align}
For the inversion-symmetric perturbation, the dispersion calculated using Eq.~\eqref{eq:mu} contains an anisotropic sDP \cite{wallbank_prb_2013, yankowitz_natphys_2012, park_prl_2008, guinea_ptrsa_2010} with Dirac velocity component $\approx 2U^+_0 /b$ in the direction of the sBZ edge and $\approx v$ in the perpendicular direction. A similar Hamiltonian to Eq.~\eqref{eq:mu} is obtained for the other two non-equivalent edges of the first sBZ, so that three of these sDPs are visible between the first and second valence miniband in Fig.~\ref{fig:parameter_space} (c). The gap at the sDP, opened by the inversion-asymmetric terms is,
\begin{align}
\Delta_{\vect \mu} &=\frac{|U^+_0 (s    U^-_1-U^-_3)+U^-_0 (s   U^+_1-U^+_3)|}      {\sqrt{{U^+_0}^2+{U^-_0}^2}}\nonumber
\end{align}
 
For the points $\vect \kappa=(\vect b_4+\vect b_5)/3$ and $-\vect \kappa$, zone folding brings together three degenerate plane wave states, $|\xi( \vect\kappa +\vect q)\rangle$, $|\xi( \vect\kappa +\vect b_1+\vect q)\rangle$, and $|\xi( \vect\kappa+\vect b_2 +\vect q)\rangle$ (where $\xi=\pm$ for the two inequivalent sBZ corners), whose splitting is determined by 
\begin{align*} %\label{eq:H_kappa}
&\op{H}_{\xi(\vect\kappa+\vect q)}\!=\! \begin{pmatrix} 
  \frac{svb}{\sqrt{3} }\!+\!s v q_x       &w_{\xi}	 & w_{\xi}^*\\
w_{\xi}^*	& \frac{svb}{\sqrt{3} }\!-\!sv\frac{q_x-\sqrt 3 q_y}{2 }     	&-w_{\xi}     \\
w_{\xi}   				& -  w_{\xi}^*         	 &\frac{svb}{\sqrt{3} }\!-\!sv\frac{q_x+\sqrt 3 q_y}{2 }\\
 \end{pmatrix}\!\!,\nonumber\\
 &w_{ \xi}\!\approx\!\frac{1}{2}\!\!\left[\! \left(\!U^+_0  \!\!-\!\!2s\xi U^+_1\!+\!\sqrt 3 \xi U^+_3\!\right)\!\!+\!i\xi\!\left(\! U^-_0 \!+\!2 s\xi U^-_1\!\!-\!\!\sqrt3\xi U^-_3 \!    \right)  \! \right]\!. 
 \end{align*}
The inversion-symmetric terms in $\hat H_{\xi(\vect\kappa+\vect q)}$ partially lift the $\xi\vect \kappa$-point degeneracy into a singlet with energy $(\frac{svb}{\sqrt 3}-2w_{\xi} )$ and a doublet with energies $(\frac{svb}{\sqrt 3}+w_{\xi} )$, so that a distinctive sDP \cite{guinea_ptrsa_2010, ortix_prb_2012,wallbank_prb_2013} characterized by Dirac velocity $v_{\vect\kappa}=\frac{v}{2}\left[1+3(\sqrt 3U^+_0-\xi U^+_3)/(2vb)\right]$ \cite{wallbank_thesis_2014} is always present at  $\pm\vect \kappa$ somewhere in the spectrum. This single isotropic sDP is visible in Fig.~\ref{fig:parameter_space} (d) at the first valence miniband edge. Once again, the inversion symmetric terms open a gap at the sDP, 
\begin{align}
\Delta_{\xi\vect\kappa} = \sqrt{3} | U^-_0 + 2s\xi U^-_1 
- \sqrt 3 \xi U^-_3|. \nonumber
\end{align}

\subsection{Microscopic models for moir\'e perturbation parameters} \label{sec:microscopic_models}
\emph{Interlayer hopping models.}
To describe the interlayer coupling in graphene/{\hbn} heterostructure, several studies \cite{kindermann_prb_2012, san-jose_prb_2014, san-jose_prb_2014_2, moon_prb_2014} made use of the similarity between this system and twisted bilayer graphene. In the latter \cite{santos_prl_2007, bistritzer_prb_2010, bistritzer_pnas_2011, santos_prb_2012}, the electronic structure can be described by a bilayer-like Hamiltonian, in which the intralayer blocks are given by the Dirac Hamiltonian and the interlayer blocks describe the modulation of the interlayer coupling as a function of the position within the mSL \cite{bistritzer_pnas_2011}. Applied to the $K$ valley of the graphene/{\hbn} heterostructure, such a Hamiltonian takes the form
\begin{align*}\begin{split}% \label{eq:H_hopping_model}
& \op{H}_{\mathrm{2layer}} = \left(\begin{array}{cc}
v\vect{\sigma}\cdot\vect{p} & \op{T}(\vect{r}) \\
\op{T}^{\dagger}(\vect{r}) & \op{H}_{\mathrm{hBN}}
\end{array}\right), \\
& \op{H}_{\mathrm{hBN}} = \left(\begin{array}{cc}\epsilon_{N}&0\\0&\epsilon_{B}\end{array}\right),\\
& \op{T}(\vect{r})=\frac{1}{3}\sum_{j=0,1,2} e^{-i (\hat R_{2\pi j/3 }\vect \kappa)\cdot \vect r} 
\begin{pmatrix}  \gamma_N & \gamma_B e^{-i \frac{2\pi }{3}j}\\  \gamma_N e^{i \frac{2\pi }{3} j } & \gamma_B \end{pmatrix}.
\end{split}\end{align*}
Above, the Dirac Hamiltonian is used for the graphene layer, $\op{H}_{\mathrm{hBN}}$ describes the {\hbn} layer with $\epsilon_{B}$ and $\epsilon_{N}$ characterising the on-site energy of the boron  and nitrogen sublattices, and $\op{T}(\vect{r})$ describes the spatially varying interlayer coupling, with $\gamma_{B}$ and $\gamma_{N}$ the hopping integrals from graphene to the boron and nitrogen sites respectively. For the energies of interest, $|\epsilon|\ll|\epsilon_N|,|\epsilon_B|$, the Hamiltonian can be projected on the Hilbert space of the graphene layer, so that the perturbation is parametrised by Eq.~\eqref{eq:H} with,
\begin{align}\label{eq:model_params} 
\{U_{i=0,1,3}^\pm\}= V^\pm \left\{\frac{\pm1}{2} , -1 ,\frac{-\sqrt 3}{2}\right\},
\end{align}
where,
\begin{align*}
 V^+=\frac{1}{18}\left(\frac{\gamma_N^2}{\epsilon_N}+\frac{\gamma_B^2}{\epsilon_B}\right),\quad 
 V^-=\frac{\sqrt 3}{18}\left(\frac{\gamma_N^2}{\epsilon_N}-\frac{\gamma_B^2}{\epsilon_B}\right). 
\end{align*}

\emph{Intralayer potential models.} 
A complementary description of the mSL perturbation for electrons in the graphene/{\hbn} heterostructure
concentrates on the significant \cite{jung_prb_2013} electrostatic potentials generated by the presence of two distinct atomic species in the underlay.
In contrast to graphene, in which every carbon atom (atomic number 6) provides one $\pi$-electron, in {\hbn} the boron atoms (atomic number 5) provide no $\pi$-electrons whereas the nitrogen atoms (atomic number 7) provides two $\pi$-electrons each. Then, the electrostatic potential in the underlay can be modelled as a trigonal lattice of $+2|\text e|$ point charges mimicking the core charges of the nitrogen atoms, compensated by a homogeneous background charge density mimicking the dispersed cloud of the $\pi$-electrons\cite{wallbank_prb_2013}.
The matrix elements of the resulting potential, taken between sublattice Bloch states $i$ and $j$ ($i,j=A\text{ or }B$), acting on 
the low energy Dirac spinors of the graphene $K$ valley, are given by the long wavelength components of  
\begin{align}\label{eq:point}
  \delta \!H_{ij}&=\frac{-2 
 \text{e}^2}{4\pi\epsilon_0}\sum_{\vect R_{N}}\int\! d\!z \frac{L^2\Phi^*_{Ki}(\vect r,z)\Phi_{Kj}(\vect r,z)}{\sqrt{(\vect r-\vect R_{N})^2+(z-d)^2}}.
\end{align}
Here, $\vect R_N$ are the positions of the nitrogen sites, $L^2$ is the total area of the graphene sheet, $3.22\text{\AA}\leq d\leq3.5\text{\AA}$ \cite{giovannetti_prb_2007} is the graphene-{\hbn} separation and $\Phi_{K,i}(\vect r,z)$ are the Bloch wavefunctions of graphene $2p_z$ orbitals exactly at the $K$ point. The long wavelength components of Eq.~ \eqref{eq:point} are extracted \cite{wallbank_prb_2013} by re-writing the Bloch wave functions using the in-plane 2D Fourier transform of the graphene $2p_z$ orbitals $\hat \psi(\vect q,z)$, and the dominance of the simplest moir\'e harmonics in the mSL potential is set by the rapid decay of $\hat \psi(\vect q,z)$ for $|\vect q|>|\vect K|$ at $z=d$ (see Eq.~\eqref{eq:pz_FT}). Interestingly, this model is parametrised by Eq.~\eqref{eq:H} except with,
\begin{align*}
&V^+\approx \frac{4\pi \text{e}^2}{ \sqrt{3} \epsilon_0 a^3}  \int\! dz  \hat\psi^*(\vect K,z)   e^{-\frac{4\pi}{\sqrt3 a}|z-d|}   \hat\psi(\vect K , z),\qquad V^-=0.
\end{align*}
The relation $V^-=0$ is prescribed by the inversion symmetry resulting from the assumption that only the nitrogen atoms are responsible for the mSL potential. Non-zero values for $V^-$ are obtained when a charge on the boron sites is included. This contribution can be obtained in a simple way from Eq.~\eqref{eq:model_params} by using the coordinate shift, Eq.~\eqref{eq:rotate}, which is then added to the part of the moir\'e potential due to the nitrogen atoms.
A similar approach is employed in Ref.~\cite{ortix_prb_2012}, obtaining a similar mSL perturbation but without the modulation of the nearest neighbour hopping between carbon atoms.\newline

\emph{Numerical models.}
Many studies of the  graphene/{\hbn} heterostructure have employed numerical models based on \emph{ab initio} and/or numerical tight-binding approaches.
One obstacle to the application of such techniques arises from the fact that no true (perfectly translationally invariant) unit cell exists for this system, due to the incommensurability between the graphene and {\hbn} lattices.
Several early \emph{ab initio} studies avoided this problem by assuming the two crystals to be lattice matched, for example by contracting the {\hbn} lattice and assuming perfect rotational alignment. The graphene in this lattice matched system typically has a strongly broken sublattice symmetry, resulting in predictions of large band gaps $\sim50\,$meV opened at the primary Dirac point \cite{giovannetti_prb_2007, slawinska_prb_2010, fan_apl_2011, kharche_nanolett_2011, sachs_prb_2011}. However, already calculations by Sachs and co-authors \cite{sachs_prb_2011} showed that the van der Waals adhesion of the graphene to the {\hbn} underlay is insufficient to cause the system to become lattice matched.

Other \emph{ab initio} models of the graphene/{\hbn} heterostructure have used a large commensurate unit cell (e.g.~55 carbon atoms on 56 {\hbn} atoms) to approximate the incommensurate moir\'e unit cell \cite{kharche_nanolett_2011,martinez-gordillo_prb_2014}.  
Also, numerical tight-binding calculations using a large commensurate unit was employed to explain some of the experimental work \cite{dean_nature_2013, hunt_science_2013}, and more recently \cite{moon_prb_2014} this approach was found to agree a continuum model similar to Eq.~\eqref{eq:H_simple}. 

A further development is detailed by Jung and co-authors in Ref.~\cite{jung_prb_2013}, with similar approaches employed in Refs.~\cite{sachs_prb_2011, bokdam_prb_2014, jung_arxiv_2014}. Using this methodology, the numerical simulation is performed using unit cells containing only four atoms in total and a forced commensuration. However, the incommensurability is then included by simulating many such unit cells with a graphene-{\hbn} coordination specific to each local area in the mSL supercell. Then the mSL perturbation can be parametrised using the dominance of the simplest mSL harmonics in the mSL perturbation. 
\newline 
 
\emph{Models incorporating spontaneous lattice deformations.}
As mentioned in the introduction, a relaxation of the atomic positions of the crystal layers in unencapsulated highly aligned graphene/{\hbn} heterostructure has been observed experimentally \cite{woods_natphys_2014}.
Here, the local displacements of the two layers results from a competition between the graphene-{\hbn} adhesion landscape and the elasticity of the graphene and {\hbn} layers. 
The former contribution varies on the atomic scale and is found to be minimised for a commensurate AB stacking in which the boron atom is situated directly beneath a carbon site, and the nitrogen directly beneath the hexagon centre \cite{giovannetti_prb_2007,fan_apl_2011,sachs_prb_2011,slotman_annphys_2014}. In contrast, the elastic contribution prefers the 
incommensurate arrangement (described by Eq.~\eqref{eq:hatM}) in which both layers maintain the rigid arrangement expected of individual layers. Similar to experimental observations \cite{woods_natphys_2014}, both the analytical \cite{san-jose_prb_2014,san-jose_prb_2014_2} and \emph{ab initio} \cite{jung_arxiv_2014,bokdam_prb_2014} models show an expansion of the regions of the mSL in which the preferred graphene-{\hbn} stacking is obtained. This is particularly pronounced at small misalignment angles ($\theta\lesssim 1^\circ$). Similar effects have also been observed in molecular dynamics simulations \cite{wijk_arxiv_20014}, with comparison to the experiment \cite{woods_natphys_2014} suggesting that the carbon-nitrogen interaction is two or three times stronger that the carbon-boron interaction. The breaking of the sublattice symmetry in the commensurate phase opens a band gap at zero-energy \cite{san-jose_prb_2014_2}, which in \emph{ab initio} models is enhanced by the electron-electron 
interaction to a value of $\Delta_0\sim20-30\,$meV \cite{jung_arxiv_2014,bokdam_prb_2014}. Also, the strain induces vector and scalar potentials in Hamiltonian \eqref{eq:H} which produce a significant non-trivial contribution to the mSL potential \cite{san-jose_prb_2014_2, neek-amal_apl_2014, neek-amal_apl_2014_2}.

\subsection{Homogeneously strained graphene in heterostructure with hBN}\label{sec:strained_heterostructures}

\begin{figure}[tbhp]
\centering
\includegraphics[width=.5\textwidth]{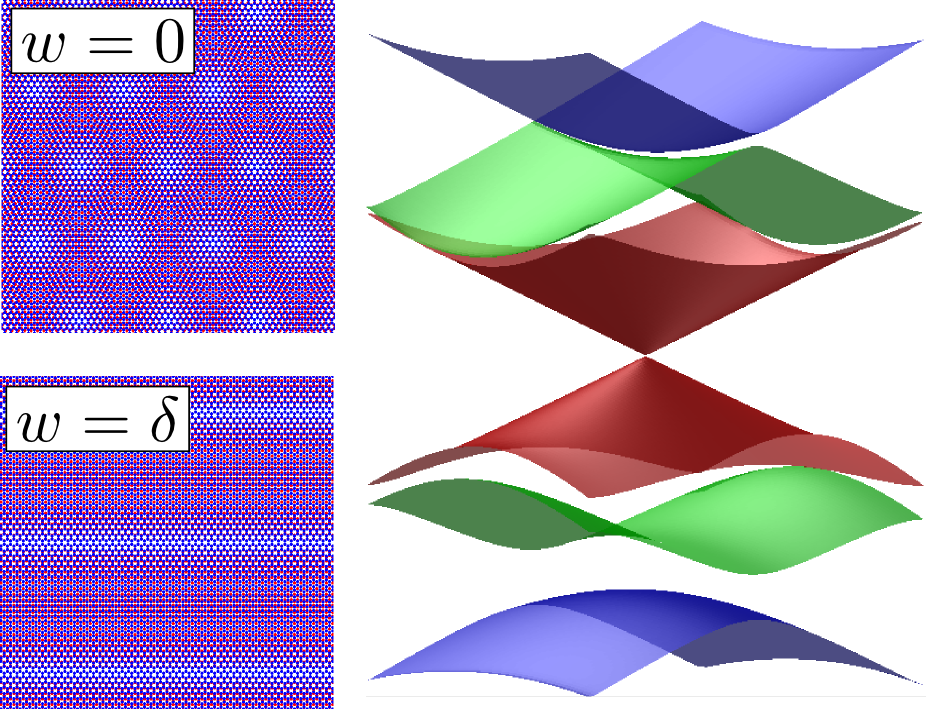}
\caption { 
Left: The moir\'e superlattice in unstrained (top) and strained (bottom) heterostructures. Right: The bandstructure of the strained heterostructure, calculated using mSL perturbation set by Eq.~\eqref{eq:model_params} with $V^+ =22\,$meV.
 }
\label{fig:strained_bands}
\end {figure}

Figure \ref{fig:strained_bands} provides a simple example of how a small strain, $w\sim \sqrt{\delta^2+\theta^2}$, (either in the graphene layer or the substrate) can induce qualitative changes in both the geometry of the mSL and its corresponding miniband structure. The top left panel shows the usual hexagonal mSL of an unstrained and perfectly aligned ($\theta=0$) heterostructure, while in the lower left panel a uniaxial strain $w=\delta$ has been applied to compensate the lattice mismatch along the $x$-axis so that the mSL becomes one-dimensional. This strain is incorporated into the Hamiltonian \eqref{eq:H} using
\begin{align*}%\label{eq:u_homogeneous}
 \vect u^{\text{Gr}}(\vect r)=\begin{pmatrix}w&0\\0&-\sigma w \end{pmatrix}\begin{pmatrix}x\\y\end{pmatrix},
\end{align*}
where we have taken $w=\delta$ and used $\sigma = 0.165$ \cite{blakslee_jap_1970} for the Poisson ratio of graphene. For this particular choice of the displacement field $\vect u^{\text{Gr}}(\vect r)$, the strain-induced vector and scalar potentials in the Hamiltonian \eqref{eq:H} correspond to a trivial momentum and energy shift of the Dirac point, and can therefore be neglected.
In this case the translational symmetry of Hamiltonian \eqref{eq:H} is described by a single reciprocal lattice vector $\vect G=\frac{2\pi \delta (1+\sigma)}{\sqrt 3 a} \hat{ \vect y}$. The corresponding band structure is then characterised by a wavevector $\vect k=(k_x,k_y)$ with $|k_y|<|\vect G|/2$ and $k_x$ unbounded.
This is shown, for a finite range of $k_x$ in the right panel of Fig.~\ref{fig:strained_bands}, and does not contain any of the sDPs characteristic of the bandstructure of the unstrained heterostructure with a similar mSL perturbation (compare with Fig.~\ref{fig:parameter_space}(d)). Interestingly, the translational symmetry of the strained one-dimensional mSL depends very sensitively on the direction of 
the applied strain, so that the corresponding spectral support develops a fractal structure as a function of the angle at which the strain is applied \cite{cosma_faraday_2014}.

\subsection{Topological aspects of the graphene/{\hbn} heterostructure}
One fundamental property of graphene is the Berry phase \cite{berry_prsla_1984} of $\pm \pi$ acquired by an electron after traversing one orbit around the Dirac point \cite{ando_jpsj_1998, novoselov_nature_2005, zhang_nature_2005}. For a general graphene/{\hbn} heterostructure with broken inversion symmetry and gapped Dirac spectrum (at either the primary or secondary Dirac point), this phase is smeared out into a non-zero Berry curvature $\Omega(\vect k)$ over a range of energies of the order of the band gap \cite{song_arxiv_2014,san-jose_prb_2014_2}. When an external electric field is applied, this Berry curvature generates an 'anomalous' contribution to the electron's velocity in the direction perpendicular to the applied field \cite{xiao_revmodphys_2010},
\begin{gather*}
 v(\vect k)=\frac{\partial \epsilon_n(\vect k)}{\partial \vect k }+e\vect E\times \vect l_z \Omega(\vect k), \\
 \Omega(\vect k)=i\left[\langle\partial_{k_x}u_n(\vect k)|\partial_{k_y}u_n(\vect k)\rangle-\langle\partial_{k_y}u_n(\vect k)|\partial_{k_x}u_n(\vect k)\rangle\right] , \nonumber
\end{gather*}
with $u_n(\vect k)=e^{-i\vect k\cdot \vect  r}\psi_{n,\vect k}(\vect r)$ the cell-periodic part of eigenstate.
 In transport experiments \cite{gorbachev_science_2014}, this is manifested as a transverse current, yielding a Hall-like conductivity for $B=0$ \cite{xiao_prl_2007, xiao_revmodphys_2010},
\begin{align*}
 \sigma_{xy}= g_se^2 \int\frac{d\vect k}{(2\pi)^2}\Omega(\vect k)f(\vect k),
\end{align*}
with $f(\vect k)$ the occupancy factor and $g_s=2$ the spin degeneracy. Due to the time-reversal symmetry, the transverse current has opposite signs for the two graphene valleys. 
However, intervalley scattering of Dirac electrons in graphene/{\hbn} heterostructures is weak.
In the transport experiment\cite{gorbachev_science_2014}, the effect was observed as a non-local voltage in a narrow energy range near primary and secondary Dirac points at finite distances from the direct current path. 

An alternative line of enquiry is motivated by the fact that, the Dirac Hamiltonian subject to a spatial varying mass term has topologically protected zero energy modes wherever the mass changes sign \cite{martin_prl_2008, semenoff_prl_2008}. For the graphene/{\hbn} heterostructure this has been applied to models \cite{kindermann_prb_2012, titov_prl_2014} in which the sublattice-asymmetric mSL potential is treated as the dominant perturbation. In this model the conducting/insulating behaviour at zero-energy is controlled by the classical percolation \cite{cardy_prl_2000} of the zero energy modes (which is achieved when the spatial average of the mass vanishes). It has also been predicted \cite{uchoa_arxiv_2014} that the electron-electron interaction leads to a spontaneous spin-valley ordering for these zero-energy modes.

\section{Optical absorption spectroscopy}
\begin{figure}[tbhp]
\centering
\includegraphics[width=.6 \textwidth]{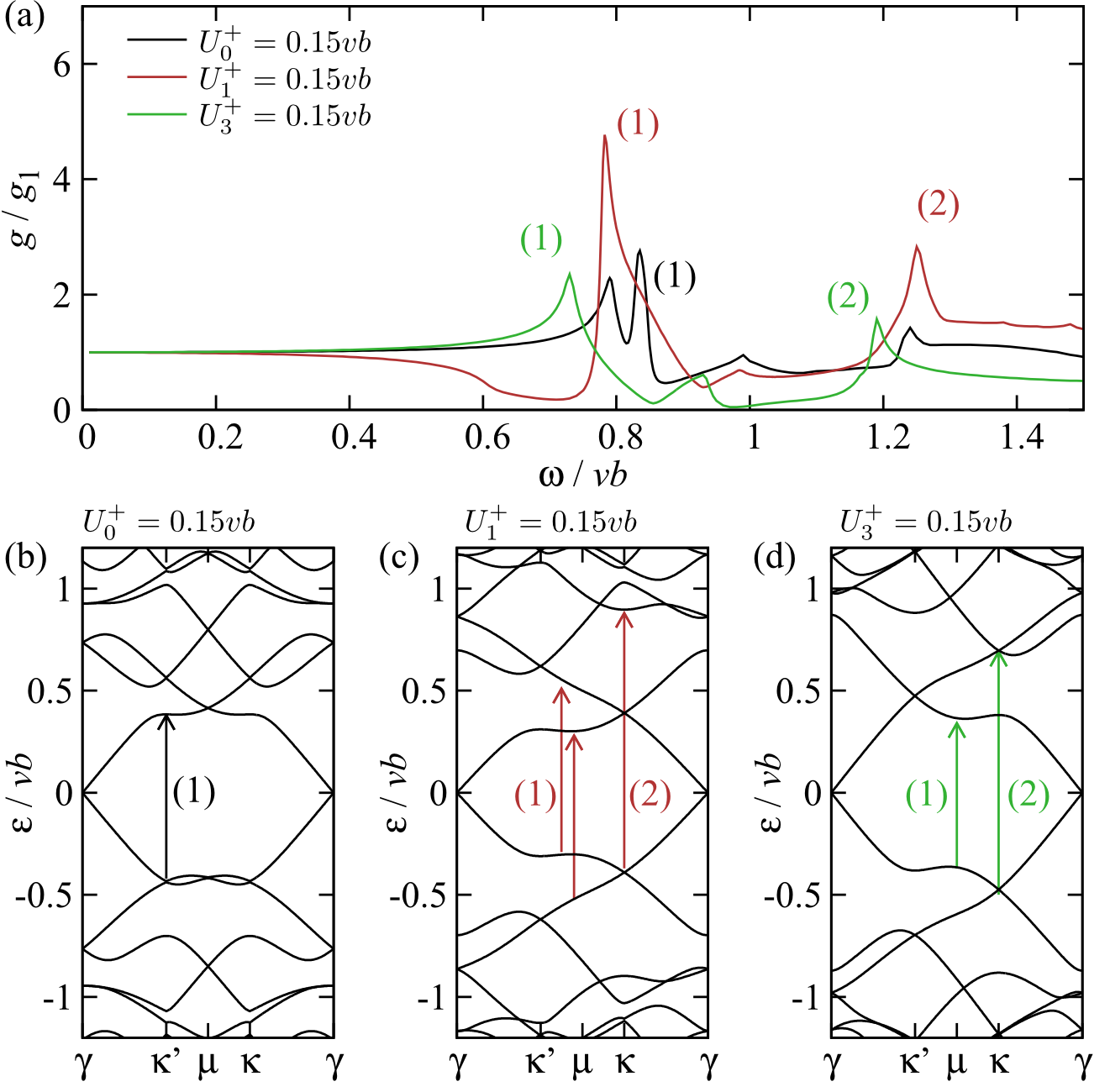}
\caption{ 	
(a) Absorption spectra for each of the inversion symmetric mSL perturbation terms for zero chemical potential. (b)--(d) The corresponding band structures with the transitions that make the strongest contribution to the labelled peaks in (a) marked with vertical arrows. Figure adapted from Ref.~\cite{abergel_njp_2013}: Abergel \emph{et.~al.}, New Journal of Physics {\bf 15}, 123009 (2013), (published under a CC BY licence).
}
\label{fig:optical_spec}
\end {figure}

In contrast to transport measurements, optical absorption spectroscopy can probe the strongly reconstructed spectrum at the edge of the first miniband of the graphene/{\hbn} heterostructure without (in principle) the necessity of first shifting the chemical potential to such energies.
The optical absorption coefficient is affected by both the changes to the energy spectrum, and the changes to the wavefunction of graphene electrons induced by the {\hbn} underlay \cite{abergel_njp_2013,shi_natphys_2014}. Similar to the studies on twisted bilayer graphene \cite{moon_prb_2013,moon_prbR_2013}, this leads to a modulation of graphene's otherwise universal optical absorption coefficient of $g_1 = \pi e^2/\hbar c\approx2.3 \% $
\cite{nair_science_2008}, which will be most pronounced in the spectral range around $\omega \sim vb$. For much lower photon frequencies, the electron states are almost the same as in the unperturbed Dirac spectrum, whereas photons of much higher energies induce transitions between numerous overlapping minibands so that individual spectral features will be indistinguishable.
Several examples of absorption spectra of an unstrained undoped graphene/{\hbn} heterostructure, calculated \cite{abergel_njp_2013} for realisations of the moir\'e perturbation in which only one of the three inversion-symmetric terms is present, are displayed in Fig.~\ref{fig:optical_spec}. Each absorption spectrum contains a characteristic pattern of peaks originating from various transitions between the minibands indicated with arrows on the corresponding bandstructure (Fig.~\ref{fig:optical_spec}(b-d)).

\begin{figure}[tbhp]
\centering
\includegraphics[width=.6 \textwidth]{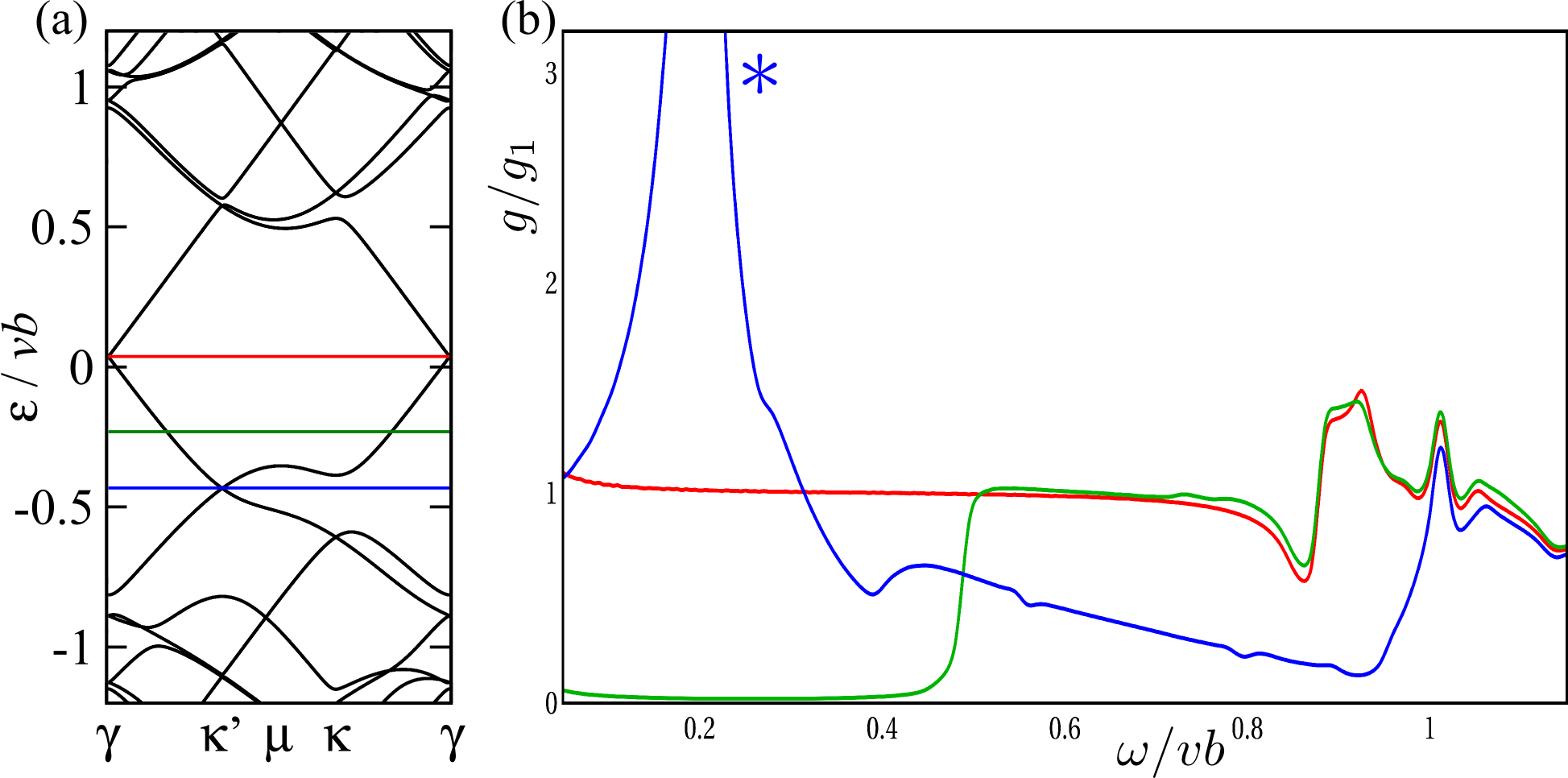}
\caption{
(a) Miniband structure, calculated using the mSL perturbation set by Eq.~\eqref{eq:model_params} with $V^+=0.063\,$vb in \eqref{eq:model_params}. (b) Corresponding absorption spectra, calculated for the various chemical potentials indicated with coloured horizontal lines in (a). 
	}
\label{fig:optical_spec_model}
\end {figure}

The experimental measurement of the optical absorption spectra of graphene/{\hbn} heterostructures is compounded by a large absorption background produced by the hBN underlay. To overcome this difficulty, Shi and co-workers \cite{shi_natphys_2014} produced an electrostatically gated heterostructure and, by comparing absorption spectra for different graphene chemical potentials, could subtract the background contribution. Interestingly, they found the measured absorption spectra consistent with that calculated using the mSL perturbation set by Eq.~\eqref{eq:model_params} which is based on the simple microscopic models.
The right panel of Fig.~\ref{fig:optical_spec_model} shows the theoretical absorption spectra, calculated using Eq.~\eqref{eq:model_params} with $V^+=0.63vb$ for several values of hole doping.
The left panel shows the miniband structure with coloured lines used to indicate the chemical potentials used in the right panel. By comparing the absorption spectra calculated for the undoped heterostructure (red line) to that in a slightly doped heterostructure (green line), it is seen that the effect of a small shift in chemical potential ($|\mu|< vb/2$) consist of the usual Pauli blocking of transitions with $\omega <2\mu$, which is the same as the effect on unperturbed graphene. In contrast, for higher hole doping (blue line, chemical potential tuned to the valence band sDP), new absorption channels can be opened. This results in the additional absorption peak marked with an asterisk in the right panel.

\section{Fractal spectrum of magnetic minibands of the graphene/{\hbn} heterostructure in a strong magnetic field}
 
The fractal spectrum of electron waves constrained to two dimensions and exposed to both an in-plane periodic potential and a strong perpendicular magnetic field \cite{zak_physrev_1964,brown_physrev_1964,hofstadter_prb_1976} is one of the most spectacular results in the quantum theory of solids \cite{stat_phys_LL}. When the magnetic flux, $\phi\equiv BS=\frac{p}{q}\phi_0$ threading each unit cell (area $S$) of the system becomes a rational fraction ($p$, $q$ integers) of the flux quantum, $\phi_0=\frac{h}{e}$, each Landau level is fractured into $p$ magnetic bands. 
Since the width of the magnetic bands only becomes appreciable when $\phi\gtrsim 0.2\phi_0$, their experimental observation at sustainable magnetic fields requires the use of systems with a large lattice constant (e.g.~$A\sim10\,$nm).

Early attempts to observe the fractal spectra focused on two-dimensional electrons in periodically patterned GaAs/AlGaAS heterostructures \cite{weiss_europhys_1989,weiss_prl_1991,pfannkuche_prb_1992,ferry_pqe_1992, albrecht_prl_1999, schlosser_prl_2001, albrecht_prl_2001, geisler_prl_2004}, where the superimposed superlattice period was made sufficiently large. However, the technological difficulties in producing such devices without introducing substantial disorder, and accessing the electron gas buried relatively deep within the sample, resulted in only partial success. In contrast, the graphene/{\hbn} heterostructure, with the high quality of the exfoliated flakes, its naturally occuring mSL, and the ease of doping of graphene by electrostatic means, provides the perfect system to observe the fractal magnetic bands.

Already, several observations of the fractal bands have been reported in transport measurements, both for encapsuted \cite{ponomarenko_nature_2013} and non-encapsulated \cite{hunt_science_2013} heterostructures, as well as for devices made of bilayer graphene on aligned {\hbn} \cite{dean_nature_2013}. 
Here, the signatures of the magnetic bands are the strongest when $\phi/\phi_0=1/q$ \cite{ponomarenko_nature_2013,dean_nature_2013, hunt_science_2013}. Upon doping with 4 holes per moir\'e unit cell (corresponding to emptying the first $B=0$ miniband), the Hall resistivity is observed to change sign for every $\phi/\phi_0=1/q$, indicating the recurrent generation of electron-like orbits in graphene's valence band due to the magnetic bands \cite{ponomarenko_nature_2013}.
Also, transport measurements \cite{hunt_science_2013}, and, more recently, capacitance measurements \cite{yu_natphys_2014} have revealed the lifting of spin and valley degeneracies of the magnetic minibands as a result of the electron-electron interaction.
There have also been several theoretical works on the magnetic bands of graphene/{\hbn} heterostructure \cite{chen_prb_2014, diez_prb_2014, moon_prb_2014, chizhova_arxiv_2014}, in addition to a plethora of works on the magnetic bands in the related twisted bilayer system \cite{bistritzer_prb_2011,moon_prb_2012,wang_nano_lett_2012,hasegawa_prb_2013,moon_prbR_2013}.

\subsection{Magnetic translational symmetry}\label{sec:mag_bands}

To include a magnetic field in Hamiltonian \eqref{eq:H_simple} it is convenient \cite{chen_prb_2014} to use a Landau gauge for the magnetic vector potential, $ \vect A\!=\!B x_1(-\vect A_1+2\vect A_2)/(\sqrt3 A)$, written using a coordinate system adapted to the hexagonal symmetry of the mSL, $ \vect r=x_1 \frac{\vect A_1}{A} +x_2 \frac{\vect A_2}{A}$. Then Hamiltonian \eqref{eq:H_simple} commutes with the magnetic translations \cite{zak_physrev_1964,brown_physrev_1964, stat_phys_LL} in the group, $G_M=\{\Theta_{\vect X}\equiv e^{ieBm_1A\frac{\sqrt3}{2}x_2} T_{\vect X}, \vect X=m_1\vect A_1+m_2\vect A_2\}$, where  $T_{\vect X}$ is a geometrical translations on the mSL, and
\begin{gather*}
\Theta_{\vect X}\Theta_{\vect X'} =   e^{i 2\pi\frac{p}{q} m_1'm_2} \Theta_{\vect X+\vect X'}, \nonumber\\
\Theta_{\vect X}\Theta_{\vect X'} =      e^{i2\pi\frac{p}{q}   (m_1'm_2-m_1m_2') }  \Theta_{\vect X'}\Theta_{\vect X}.\label{eq:mag_tran}
\end{gather*} 
The subgroup of $G_M$ made of translations $ \vect R=m_1q\vect{A}_1+m_2q\vect{A}_2 $ on a $(q\times q)$-enlarged superlattice is isomorphic to the simple group of translations, $T_{\vect R}$, so that its eigenstates, $ \Theta_{\vect R}|\Phi_{t, \vect k}^{n,j}\rangle =e^{i \vect k\cdot\vect R} |\Phi_{t, \vect k}^{n,j}\rangle$, form a plane wave basis with a magnetic Brillouin zone with area $q^2$-times smaller area than the mSL Brillouin zone. Moreover, since the non-abelian group $G_M$ has $q$-dimensional irreducible representations \cite{brown_physrev_1964}, the spectrum of such plane-wave states is $q$-fold degenerate.

To calculate the magnetic minibands, we project the Hamiltonian \eqref{eq:H_simple} on to the basis of Bloch functions, $|\Phi_{t, \vect k}^{n,j}\rangle$, built from the wave functions of the unperturbed Landau levels, $\psi_n^{ k_2}(\vect r)$,
\begin{align} \label{eq:bloch_wf}  
&|\Phi_{t, \vect k}^{n,j}\rangle =\!
\frac{1}{\sqrt N}\!\sum_{r} e^{-ik_1 qAr}\psi_n^{ k_2+\frac{\sqrt3}{2}b(pr+j+\frac{tp}{q})},\quad j=1,\cdots p,  \\
&\psi_{0}^{k_2}\!=\! \frac{e^{ik_2 x_2}}{ \sqrt L}\!\colvect{\frac{1-\beta}{2}\varphi_0}{\frac{1+\beta}{2}\varphi _0}\! ,\quad
\psi_{n\neq0}^{k_2}\!=\!  \frac{e^{ik_2 x_2}}{\sqrt{2L} }\!\colvect{\varphi _{n^{\!-} }}{\text{sign}(n)  \beta e^{i\frac{2\pi}{3}} \varphi _{ n^{\!+}  }}\!, \nonumber\\
&\varphi_n \!=\!     A_n   e^{-\frac{z^2}{2}+\beta \frac{i z^2}{2 \sqrt{3}}}\mathbb{H}_n(\!z\!),\quad z\!=\frac{\sqrt{3} x_1}{2\lambda_B}+\beta k_2\lambda_B. \nonumber
\end{align}
Here, $\beta\!=\!B/|B|$, $A_n\!=\!\sqrt{ 3  }/(\sqrt{n!2^{(n+1)} \lambda_B\sqrt{\pi}} )$, $n^\pm\!=\!|n|\pm(\beta\mp1)/2$,
$\mathbb{H}_n$ is the Hermite polynomial, $\lambda_B\!=\!1/\sqrt{|eB|}$  is the  magnetic length, and in the first equality the sum runs over $r=-N/2, \cdots ,N/2$,  ($N\to\infty$).
This basis is similar to the set of Bloch states for a one-dimensional chain with $p$ sites per elementary unit cell, and multiple atomic orbitals on each site labeled by the Landau level index, $n$. An addition index, $t\!=\!0,\cdots q\!-\!1$, takes into account the above-mentioned $q$-fold degeneracy. Also, $\vect k\!=\!k_1 \hat{\vect k}_1+k_2 \hat{\vect k}_2$, $|k_i|\!<\!\frac{\sqrt3}{4q}b$, $ \hat{\vect k}_i\cdot\frac{\vect A_j}{A}=\delta_{i,j}$. Since the above basis has the convenient property,
\begin{align*}
 \langle\Phi_{t, \vect k}^{n,j}| \hat H |\Phi_{\tilde t, \tilde{\vect k}}^{\tilde n,\tilde j}\rangle\sim \delta_{t,\tilde t}\delta(\vect k-\tilde{\vect k}),
\end{align*}
it is sufficient to fix $\vect k$ and consider $ t=0$ only. Also, the  $|\Phi_{t, \vect k}^{n,j}\rangle$ diagonalize the Dirac part of Eq.~\eqref{eq:H_simple} giving the energy of an unperturbed Landau level, $E_{n}= \text{sign}(n)v \lambda_B^{-1}\sqrt{2|n|}$, so that the mSL potentials can be treated perturbatively and the resulting magnetic band-structure converges when only a finite range of $n$ is included in the basis.

\begin{figure*}[t]
\center
\includegraphics[width=1\textwidth]{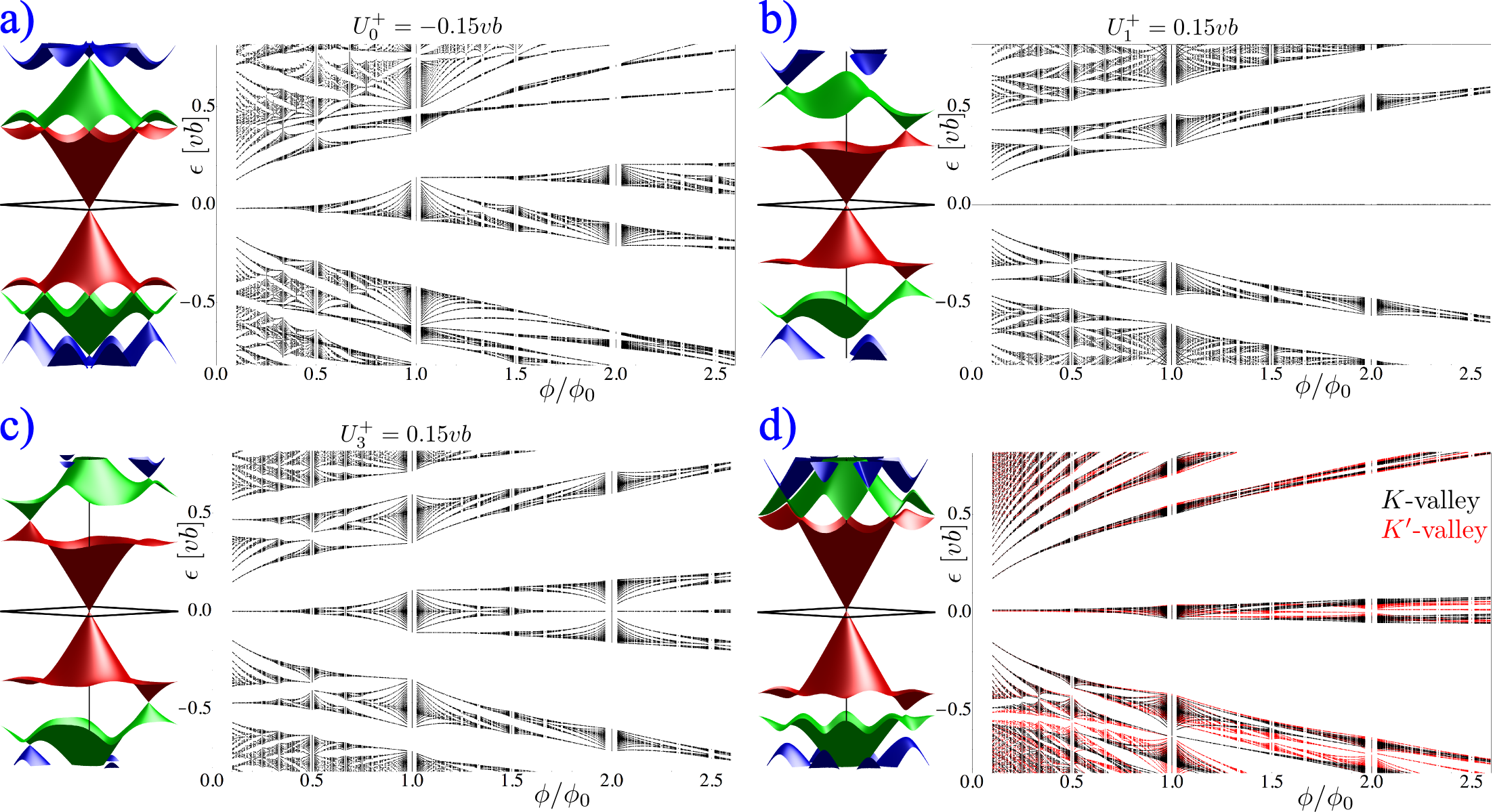} 
\caption{Zero magnetic field band structure and the corresponding support of the magnetic minibands. Panel (d) was calculated using 
$U^+_{i=0,1,3}=\left\{-0.61,-2.05,-1.32  \right\}\,$meV, $U^-_{i=0,1,3}=\left\{10.11,11.15, 8.91  \right\}\,$meV, chosen to mimic the parameter set calculated in Ref.~\cite{jung_prb_2013}. 
}
\label{fig:magnetic_miniband_spectra}
\end{figure*}

\subsection{Generic features of the magnetic miniband spectra}
Magnetic miniband spectra calculated for various mSL perturbations are displayed in Fig.~\ref{fig:magnetic_miniband_spectra}.  
In each case the magnetic miniband spectra for weak magnetic fields, $\phi\lesssim0.2\phi_0$, can be traced to a weakly broadened sequence of Landau levels associated with both the primary Dirac point as well as the sDPs. In higher magnetic fields, each Landau level is fractured to reveal the fractal magnetic miniband spectrum. 

One feature of the magnetic minibands is their tendency form a weakly gapped Dirac-like spectra. An example of this for $\phi/\phi_0=1/2$ is 
shown in the right panel of the inset in Fig.~\ref{fig:parameter_space} (d). 
The lower portion of this inset shows a colour map of the Berry curvature \cite{berry_prsla_1984, xiao_revmodphys_2010} for the lower magnetic miniband. This is concentrated at the tip of the gapped Dirac point, manifesting the topological character of such features. 
The left panel of the inset shows the magnetic miniband spectra (black points), at small deviations of the magnetic field from  $\phi=\phi_0/2$. This is grouped around \cite{chang_prl_1995} the Landau levels (blue lines) of the Dirac-like magnetic bandstructure calculated for $\phi=\phi_0/2$ (albeit with the additional $q$-fold degeneracy due to the magnetic translational symmetry, and a finite gradient of the zeroth Landau level due to the finite magnetic momentum \cite{xiao_revmodphys_2010} of the magnetic bands).  
Similar features, identified in all panels of Fig.~\ref{fig:magnetic_miniband_spectra}, are particularly pronounced in Fig.~\ref{fig:magnetic_miniband_spectra} (c). Here the electron-hole symmetry $H(-\vect r)=-H(\vect r)$, present in Hamiltonian \eqref{eq:H_simple} for the perturbation $U^+_3=0.15vb$, prescribes a gapless Dirac point at zero energy whenever $p$ is even \cite{footnote:mag_miniband_degeneracies} (clearly visible for e.g.~$\phi/\phi_0=2/1$).
A separate study by Diez and co-workers \cite{diez_prb_2014} also reveals that the generically gapped Dirac-like features, can become gapless upon varying a single mSL parameter (e.g.~the misalignment angle $\theta$).

The remaining panels of Fig.~\ref{fig:magnetic_miniband_spectra} exemplify other interesting features in the magnetic miniband spectra.
The principle novelty in Fig.~\ref{fig:magnetic_miniband_spectra} (b), calculated using the mSL perturbation $U^+_1=0.15vb$, is that it contains a completely unperturbed ``$n=0$'' Landau level traced to the main Dirac point. This is a consequence of the  electron-hole symmetry, $\sigma_z H(\vect r) \sigma_z=-H(\vect r)$, which is present for this particular mSL perturbation \cite{footnote:mag_miniband_degeneracies}.
Fig.~\ref{fig:magnetic_miniband_spectra} (a) is calculated for the mSL perturbation $U^+_0=0.15vb$. Here the zero-field spectra contains a triplet of anisotropic sDPs on the edge of the first valence miniband, in contrast to Figs.~\ref{fig:magnetic_miniband_spectra} (b-d) which display only a single isotropic sDP. Consequently, the three ``$n=0$'' Landau levels traced to these three sDPs are hybridised by the mSL perturbation so that they undergo a magnetic breakdown at a lower magnetic field ($\phi\sim0.1\phi_0$) than similar features in the other three plots ($\phi\sim0.25\phi_0$). 
Finally, the spectra displayed in Fig.~\ref{fig:magnetic_miniband_spectra} (d) was calculated for a mSL perturbation which contains a large inversion asymmetric component. In this case, the combination of spatial and time-inversion asymmetry lifts the valley degeneracy \cite{chen_prb_2014,moon_prb_2014}, so that the spectral supports for the two valleys must be shown separately (black for $K$ and red for $K'$).

\subsection{Interplay between magnetic bands and the electron-electron interaction}

It is known from graphene devices manufactured on non-aligned {\hbn} \cite{dean_naturenano_2010, dean_natphys_2011, young_natphys_2012}, suspended devices \cite{du_nature_2009,bolotin_nature_2009}, and SiO$_2$ \cite{zhang_prl_2006,jiang_prl_2007,checkelsky_prl_2008}, that the electron-electron interaction can lift the four-fold spin-valley degeneracy of each Landau levels. This effect, known as quantum Hall ferromagnetism (QHFM) \cite{nomura_prl_2006}, is usually manifested through the appearance of gaps at all integer filling factors, $\nu=\rho \phi_0/B$ ($\rho$ the carrier density), and is particularly pronounced for $\nu=0,\pm1$, corresponding to lifting the spin-valley degeneracy within the $n=0$ Landau level.

Related effects have also been observed for the magnetic minibands of the aligned graphene/hBN heterostructure, both in transport measurements \cite{hunt_science_2013} and capacitance measurements \cite{yu_natphys_2014}.
In the latter, incompressible states were traced to the band gaps created by the Landau level sequences of both the primary and secondary Dirac point. Other gaps were associated with the Landau quantization of magnetic minibands near-rational values of the flux. Further sequences of gaps could not be explained using the single-particle picture, and appeared due to the interplay \cite{apalkov_prl_2013} between the electron-electron interaction and the fractal magnetic miniband. Of these, the filling factors related to the QHFM expected in the absence of a mSL were suppressed for all rational fluxes $\phi=p\phi_0/q$ for which the band width of the magnetic minibands is large. This effect was found to be most pronounced around $\phi=\phi_0$ where the gaps at filling factors $\nu=\pm1$ (counted from charge neutrality) 
disappeared. Moreover, the Landau level sequences traced to small deviations in the magnetic field $\delta\!B$ from $\phi=\phi_0$  exhibited their own QHFM, displaying lifting of the spin-valley degeneracy with gaps set by the coulomb energy $E(\delta\!B )\sim e^2/(\epsilon \lambda_{\delta\!B })$ (here $\epsilon$ is the electrical permittivity). Also, the filling factors $|\nu|=3,4,5$ were suppressed near $\phi=\phi_0/2$.

\section{Moir\'e magnification of defects in graphene heterostructures} \label{sec:moire_magnification}

\begin{figure}[tbp]
\centering
\includegraphics[width=.65 \textwidth]{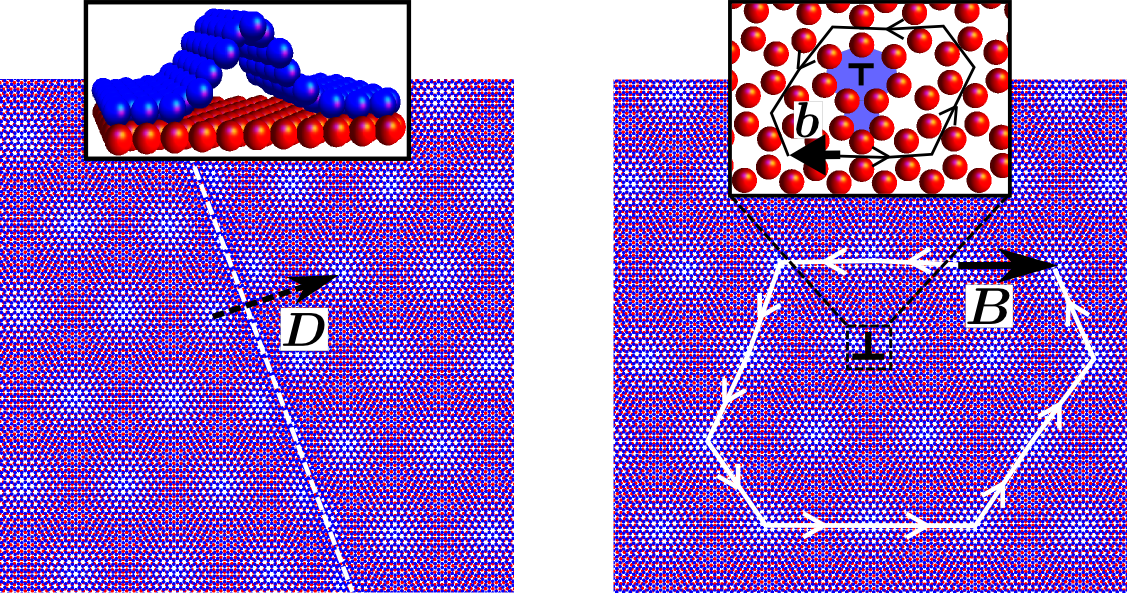}
\caption{ The mSL in an aligned graphene/hBN heterostructure with either, (a) a wrinkle in the graphene layer, or (b) a dislocation in the hBN layer. }
\label{fig:Fig_defects}
\end {figure}

Section \ref{sec:strained_heterostructures} discussed the effects of small homogeneous strains on the graphene/hBN heterostructures and its corresponding miniband structure. 
Here we use hBN as an example of a hexagonal substrate for graphene to discuss the geometrical magnification effect in which general extended or topological defects in either crystal layer generate a similar defect in the mSL, magnified by a factor $M\sim[\delta^2+\theta^2]^{-1/2}$ \cite{cosma_faraday_2014}. 
 
To track the effect of these defects on the geometry of the mSL, we follow the coordinate dependent phase factor in Hamiltonian \eqref{eq:H}, so that we identify the formal ``sites'', of the mSL as points with a constant phase, $\vect g_m \cdot \vect u(\vect R)=2\pi N$. 
Then, comparing the location of these sites with ($\vect R_d$) and without ($\vect R$) the deformation due to a defect, we define the  displacement field in the mSL by $\vect U_d\left(\vect R_d\right)\!\equiv\!\vect R_d-\vect R$, and find that
\begin{equation}
\vect U_d\left(\vect r\right) = \hat{M}\left(\vect u^{\text{Gr}}(\vect r)-\vect u^{\text{{\hbn}}}(\vect r)\right),
 \label{eq:defect}
\end{equation}
where $\hat{M}$ is the ``magnifying matrix'' given in Eq.~\eqref{eq:hatM}.

To demonstrate the use of Eq.~\eqref{eq:defect}, the left panel of Fig.~\ref{fig:Fig_defects} displays the effect of a wrinkle on the mSL. Here, the small shift
\begin{align}
\vect d\equiv\!\int_{\vect D}\!  d\!s \frac{d\vect u^{\text{Gr}}(\vect r(s))}{d\!s},\nonumber
\end{align}
in the atomic positions of carbon atoms, with respect to their position in flat graphene, across the wrinkle is reflected in a magnified shift of sites in the mSL,
\begin{gather}
\vect D =\hat {M}\vect d.
\label{eq:wrinkles}
\end{gather}
Equation \eqref{eq:wrinkles} can also be applied if the discontinuity in the mSL is caused by a step edge in the underlay~\cite{coraux_nano_lett_2008}. In this case $\vect d$ would include both the effect of curving the graphene flake over the step edge, as well as the shift in stacking characterised by the different atomic layers in the underlay.
Similarly, in the right panel in Fig~\ref{fig:Fig_defects}, a dislocation in the substrate, characterised by Burgers vector $ \vect{b} \equiv \oint d\!s \frac{d\vect u^{\text{{\hbn}}}(\vect r(s))}{ds}$ is reflected in a dislocation in the moir\'e superlattice, with the magnified burgers vector $\vect B$,
\begin{gather*}
  \vect B=-\hat {M}\vect{b} , \qquad \vect B\equiv \oint d\!s \frac{d\vect{U}_d(\vect{r}(s))}{ds}.
\end{gather*}

 \section{Conclusion}

The development of van der Waals heterostructures, created from stacks of two dimensional materials, is motivated by the possibility of creating devices in which the stacking, alignment, or interaction of the individual crystal layers adds new functionality to the device or opens a window on novel physical phenomena.
For the graphene/hBN heterostructure the hBN underlay is not just an excellent substrate for graphene, contributing to the high electronic quality of these devices. Rather, it generates a large quasi-periodic moir\'e superlattice and provides a source of Bragg scattering for the Dirac electrons in graphene.
This in turn creates a specific miniband structure featuring gaps \cite{hunt_science_2013} and secondary Dirac points \cite{yankowitz_natphys_2012}, which depends sensitively on the misalignment angle between the two crystal layers, and also appears to depend on whether the graphene is encapsulated or not with additional misaligned hBN layers \cite{woods_natphys_2014}.
The high quality of the two materials and the nanometre-scale period of the moir\'e superlattice also enabled the first observation of the fractal magnetic miniband structure \cite{ponomarenko_nature_2013}, and the subsequent lifting of the spin and valley degeneracies of the magnetic minibands by the electron-electron interaction \cite{hunt_science_2013, yu_natphys_2014}.
Further studies have observed the modulation of graphene's otherwise flat optical absorption spectra \cite{shi_natphys_2014}; the topological currents generated by regions of finite Berry curvature with in the minibands \cite{gorbachev_science_2014}; the generation of spontaneous strains in the graphene lattice \cite{eckmann_nanolett_2013}; and of course images of the moir\'e superlattice itself \cite{xue_natmat_2011}.
However all this is perhaps no more than just the tip of the iceberg, and we expect many more discoveries waiting for the graphene/hBN heterostructure. 
Moreover, there is now a growing library of two dimensional material \cite{geim_nat_perspective_2013}, and we hope the techniques developed in the study of this particular heterostructure will aid and inspire the development of many further devices created using this new diverse range of available materials.

\section*{Acknowledgements}
We would like to thank A.~Geim, P.~Kim, A.~MacDonald, F.~Guinea, A.~H.~Castro Neto, and S.~Adam, for useful discussions.
We acknowledge support from the EC Graphene Flagship Project CNECT-ICT-604391, ERC Synergy Grant Hetero2D and EPSRC EP/L013010/1.
MM-K would also like to thank the Graphene Research Centre for hospitality.


\begin{thebibliography}{999}
\bibitem{footnote:mag_miniband_degeneracies}
Use the quoted symmetry of Hamiltonian \eqref{eq:H_simple} to calculate relationships between the matrix elements of the Hamiltonian in the basis of the Bloch functions \eqref{eq:bloch_wf}. Then a lower bound can be placed on the the nullity of the corresponding Heisenberg matrix predicting zero energy eigenvalues.
\end{thebibliography}
\end{document}